\documentclass[aps,pra,twocolumn,superscriptaddress]{revtex4-1}
\usepackage{blindtext}
\usepackage{amsmath,amssymb,mathrsfs}
\usepackage{natbib}
\usepackage{subfigure}
\usepackage{tabularx}
\usepackage{epsfig}
\usepackage{longtable}
\usepackage{amsfonts}
\usepackage{rotating}
\usepackage{subfigure}
\usepackage{amsmath}
\usepackage{comment}
\usepackage{bbold} 
\usepackage{color}

\usepackage[unicode=false,bookmarks=true,bookmarksnumbered=false,bookmarksopen=false,breaklinks=false,pdfborder={0 0 1},backref=false,colorlinks=true]{hyperref}

\hypersetup{linkcolor=magenta,urlcolor=blue,citecolor=blue,pdfstartview={FitH},hyperfootnotes=false,unicode=true}

\def\be{\begin{equation}}
\def\ee{\end{equation}}
\def\bea{\begin{eqnarray}}
\def\eea{\end{eqnarray}}

\begin{document}
\title{Quantum Adiabatic Doping for Atomic Fermi-Hubbard Quantum Simulations}
\author{Jue Nan}
\affiliation{Hefei National Laboratory for Physical Sciences at
Microscale and Department of Modern Physics, University of Science and Technology of China, Hefei, Anhui 230026, China.}
\affiliation{State Key Laboratory of Surface Physics, Institute of Nanoelectronics and Quantum Computing, and Department of Physics, Fudan University, Shanghai 200433, China}

\author{Jian Lin}
\affiliation{State Key Laboratory of Surface Physics, Institute of Nanoelectronics and Quantum Computing, and Department of Physics, Fudan University, Shanghai 200433, China}

\author{Yuchen Luo}
\affiliation{State Key Laboratory of Surface Physics, Institute of Nanoelectronics and Quantum Computing, and Department of Physics, Fudan University, Shanghai 200433, China}

\author{Bo Zhao}
\email{bozhao@ustc.edu.cn}
\affiliation{Hefei National Laboratory for Physical Sciences at
Microscale and Department of Modern Physics, University of Science and Technology of China, Hefei, Anhui 230026, China.}
\affiliation{Shanghai Branch,CAS Center for Excellence and Synergetic Innovation Center in Quantum Information and Quantum Physics, University of Science and Technology of China, Shanghai 201315, China.}

\author{Xiaopeng Li}
\email{xiaopeng\_li@fudan.edu.cn}
\affiliation{State Key Laboratory of Surface Physics, Institute of Nanoelectronics and Quantum Computing, and Department of Physics, Fudan University, Shanghai 200433, China}
\affiliation{Shanghai Qi Zhi Institute, Xuhui District, Shanghai, 200032, China.}

\begin{abstract}
There have been considerable research efforts devoted to quantum simulations of Fermi-Hubbard model with ultracold atoms loaded in optical lattices. 
In such experiments, the antiferromagnetically ordered quantum state has been achieved at half filling in recent years. 
The atomic lattice away from half filling is expected to host d-wave superconductivity, but its low temperature phases have not been reached. 
 In a recent work \cite{lin2019quantum}, we proposed an approach of incommensurate quantum adiabatic doping, using quantum adiabatic evolution of an incommensurate lattice for preparation of the highly correlated many-body ground state of the doped Fermi-Hubbard model starting from a unit-filling band insulator.  Its feasibility has been demonstrated with numerical simulations of the adiabatic preparation for certain incommensurate particle-doping fractions, where the major problem to circumvent is the atomic localization in the incommensurate lattice. 
 Here we carry out a systematic study of the quantum adiabatic doping for a wide range of doping fractions from particle-doping to hole-doping, including both commensurate and incommensurate cases. 
We find that there is still a  localization-like slowing-down problem at commensurate fillings, and that it becomes  less  harmful in the hole-doped regime. With interactions, the adiabatic preparation is found to be more efficient for that interaction effect destabilizes localization.  For both free and interacting cases, we find the adiabatic doping has better performance in the hole-doped regime than the particle-doped regime. 
We also study adiabatic doping starting from the half-filling Mott insulator, which is found to be more efficient for certain filling fractions. 
\end{abstract}

\maketitle

\section{Introduction}
 
Ultracold  atoms  in optical lattices provide a fascinating platform for  quantum simulations of correlated many-body physics~\cite{bloch2008many,li2016physics,lewenstein2007ultracold,bloch2018quantum}.   
Since the atomic tunneling and interactions are both controllable in these systems, they  have  widely been used to study quantum many-body phases and quantum phase transitions.    
One main theme of quantum simulation with optical lattices is to investigate the low-temperature phase diagram of the Fermi-Hubbard model \cite{jaksch1998,hofstetter2002high,dutta2015non}, and help uncover the fundamental mechanism of high-temperature superconductivity~\cite{lee2006doping}.

Whether and how the d-wave superconductivity arises in the doped region including both hole- and particle doped cases, in the repulsive Fermi-Hubbard model has been attracting continuous research efforts~\cite{lee2006doping,2012_Metzner_RMP,2019_Jiang_Science,2020_Qin_PRX}, but this remains an open question with no consensus reached~\cite{2019_Jiang_Science,2020_Qin_PRX}, one reason being that the numerical simulations on classical computers meet fundamental challenges of exponentially growing Hilbert space of the quantum many-body system.  This makes quantum simulations of the doped Fermi-Hubbard model very much demanded as it is has potential to address the  important question of existence of d-wave superconductivity in the model.  
 With the development of ultracold atom experiments, the low-temperature antiferromagnetic phase has now been reached at half filling \cite{greif2013short,hart_observation_2015,mazurenko_cold-atom_2017}. 
The doping of antiferromagnet with hole has been realized by reducing the density of the trapped gases \cite{mazurenko_cold-atom_2017}. 
However, it is difficult to cool down the system to a sufficiently low temperature to enter the d-wave superconducting phase. 
How to optimally perform doping for the atomic Fermi-Hubbard optical lattice system while maintaining low entropy of the system demands more theoretical study.

A plausible method to maintain the quantum simulator at low entropy is to  perform the quantum adiabatic doping. 
The adiabatic quantum state preparation has been widely applied in the quantum state engineering and quantum simulation \cite{zoller2006,sorensen2010adiabatic,lubasch2011adiabatic,Duan2013,Chiu2008,lin2019quantum,sun2020realization}. 
For the fermions confined in the optical lattice, a different filling factor is achieved by adiabatically converting the lattice with one spatial period to a lattice with a different period. 
Going through an adiabatic evolution from insulating states to the doped regime, the system remains at the ground state of the instantaneous Hamiltonian.  
This protocol has been studied to prepare Fermi-Hubbard antiferromagnet insulating state\cite{lubasch2011adiabatic,Chiu2008}, and also doped ground state~\cite{lin2019quantum} with incommensurate fillings. 
For the incommensurate case, it has been found that the major difficulty in carrying out the quantum adiabatic doping is from fermion localization~\cite{lin2019quantum}. 
The physics of fermion localization occurring in the intermediate dynamics prevents efficient state preparation, causing a problem of localization slowing down. 
As a generic solution of the localization slowing down \cite{basko2006metal,oganesyan2007localization,pal_many-body_2010,kondov2015}, the atomic interaction is introduced and found to improve the preparation efficiency.

Since the focus of Ref.~\onlinecite{lin2019quantum} is to show the feasibility of the quantum adiabatic doping for incommensurate filling, 
only one filling factor at the particle-doped regime was studied. It is worth more systematic study how the quantum adiabatic doping behaves from different fillings, and in particular whether it remains efficient in the hole-doped regime should be addressed.

In this work we carry out a systematic study on the quantum adiabatic doping in a one-dimensional optical lattice for a broad range of filling factors, with numerical simulations based on the time-dependent density matrix renormalization group (DMRG) method. 
We find that the localization slowing down is a generic problem for both commensurate and incommensurate fillings. 
For the incommensurate case, the localization problem is more fundamental because the localization persists in the thermodynamic limit.
This problem also causes slowing down for the commensurate case considering a finite-size system with a localization length significantly smaller than  the system size, although it scales linearly with the system size in the thermodynamic limit. 
For the particle doping, we show that the adiabatic preparation efficiency can be enhanced by introducing atomic interaction for both commensurate and incommensurate fillings. 
For the hole doping, we find that the localization is much weaker, which we attribute to the large particle tunneling of the final lattice. The quantum adiabatic hole doping is consequently more efficient than the particle doping, and the efficiency can be further improved by including strong atomic interaction. 
Besides starting from the band insulator, we also consider the adiabatic particle doping starting from the Mott insulator at half filling. 
Our numerical simulation  shows that the quantum adiabatic doping starting form the Mott insulator has better performance for certain fillings. 
We expect these numerical results on the quantum adiabatic doping for a one-dimensional optical lattice would also shed light on the two-dimensional lattice.

\section{Theoretical Setup}

The atomic quantum simulator of the Fermi-Hubbard model consists of two-component Fermionic atoms confined in the periodical optical lattice.   
The system is described by the Hamiltonian
\begin{eqnarray}
H &=& \int d^d \mathbf{x} \psi^{\dagger}_{\sigma} \Big{(} -\frac{\hbar^2\nabla^2}{2M} +V(\mathbf{x}) -\mu \Big{)} \psi_{\sigma} \nonumber\\
&&+g \psi^{\dag}_{\uparrow}\psi^{\dag}_{\downarrow}
\psi_{\downarrow}\psi_{\uparrow}, 
\label{eq:Hamiltonian-field}
\end{eqnarray}
where $\psi_{\sigma = \uparrow, \downarrow} ({\bf x})$ is the quantum field operator for corresponding pseudospin (hyperfine state) $\uparrow$ and $\downarrow$ components,
$M$ the atomic mass, $\mu$ the chemical potential, $g$ the interaction strength between the two components, and $V(\mathbf{x})$ the optical lattice potential. We neglect the harmonic potential in our calculation for simplicity.  
In this work we first consider a one-dimensional($d=1$) band insulator as the initial state of the adiabatic evolution. In optical lattice experiments, band insulators with low entropy have been achieved~\cite{Chiu2008}. 
The initial lattice potential reads as 
\be
V_I(x) = V \mathrm{cos}(2\pi x/\lambda). 
\label{eq:initial potential}
\ee 
with $\lambda$ the lattice constant and $V$ the strength. 
Then we adiabatically convert the lattice to another one with a different period
\be
V_F(x) = V' \mathrm{cos}(2\pi x/\lambda'). 
\label{eq:final potential}
\ee 
During the time evolution, the potential has a time dependent form
\be
V(x,t) = (1-s(t/T))V_I(x)+s(t/T)V_F(x), 
\label{eq:evolutional potential}
\ee
which is standard in the context of adiabatic algorithm with $s(t/T)$ giving the path of the evolution and T the total evolution time \cite{albash2018adiabatic}.  
In the calculation, we focus at the parameter choice of $V = V'$, and examine the consequence of varying the overall lattice strength. 
Assuming that the atomic loss during the evolution is negligible, which holds when the evolution time $T$ is smaller than the lifetime of the cold atom system, the filling factor of the final state is $f=\lambda'/\lambda$. 
By controlling the the ratio between the lattice constants, a generic filling factor is accessible with the quantum adiabatic doping.

With the incommensurate potential in Eq.~\eqref{eq:evolutional potential}, the quantum dynamics  during the adiabatic evolution cannot be described by a valid tight-binding model. We thus 
have to take into account the continuous degrees of freedom of the lattice. 
In the numerical simulation, we discretize the space as $x \to j \times a$ , with $a$ the length of a grid, and $j$ the discrete index.    
This leads to a Hamiltonian
\be
 H = \sum_j \left \{ -t \left[ c_{j\sigma}^\dag c_{j+1,\sigma}  + H.c. \right] 
+ V_j c_{j\sigma}^\dag c_{j\sigma} +  Un_{j\uparrow} n_{j\downarrow} \right \}
\label{eq:Hamiltonian-discrete}  
\ee
with parameters $t=\hbar^2/(2Ma^2)$, $V_j=V(ja)$ and $U=g/a$. 
In the following calculation, we divide each spatial period of the initial lattice into $20$ grids if not specified.

The linear path $s(t/T) = t/T$ is adopted for the adiabatic evolution.  
To evaluate the performance of the adiabatic doping, we calculate the wave function overlap 
\be
\text{Overlap} = |\langle \Psi_g| \Psi_{T}\rangle|
\label{eq:overlap} 
\ee
and the excitation energy
\be
\Delta E = \langle \Psi(T) |H_F|\Psi(T) \rangle - \langle \Psi_g |H_F|\Psi_g \rangle,  
\ee 
where $|\Psi_g\rangle $ is the ground state of the final Hamiltonian $H_F$,  and $| \Psi(T) \rangle $  is the final state of the quantum adiabatic evolution. 
To confirm our finding is generic for finite-size system, we consider more than one system sizes in following numerical study.

\section{Particle doping}
 
\subsection{Free fermion}

\begin{figure}
\begin{center}
\includegraphics[width=1.0\linewidth]{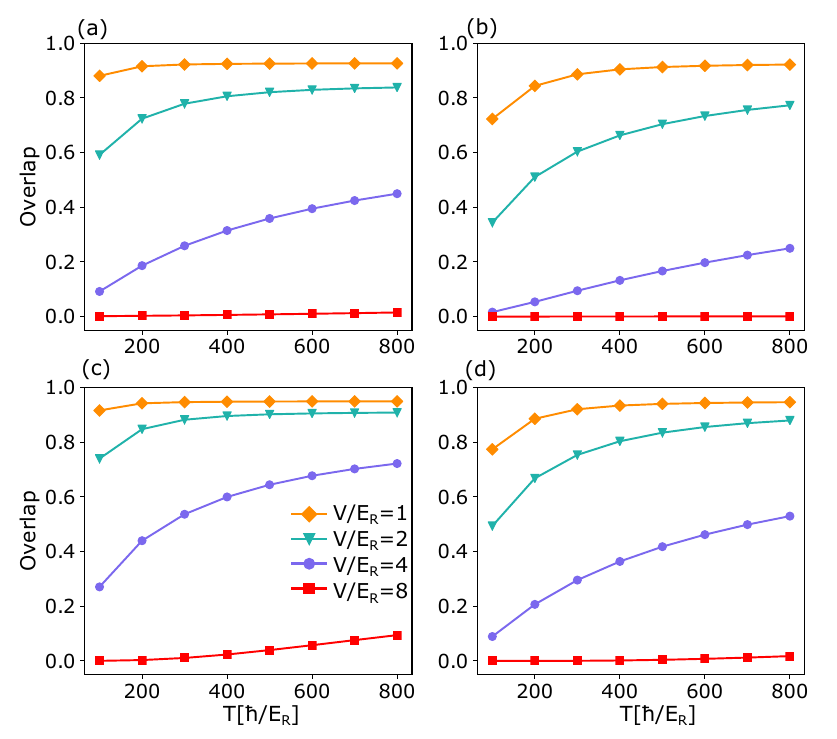}
\end{center}
\caption{
Quantum adiabatic particle doping of free fermions in one-dimensional optical lattice with rational filling factors. 
The lines show the dependence of the final state wave function overlap on total evolution time $T$ with varying strengths of lattice potential. 
The filling factor after the lattice conversion takes the rational values $f=3/4$ in (a),(b) and $f=2/3$ in (c),(d). 
For each filling factor, we consider two different system sizes, $L=36$ and $60$, with the results shown in (a),(c) and (b),(d), respectively. 
The wave function overlap systematically increases with the adiabatic time $T$. 
For weak lattice potential, the overlap approaches to $1$ with sufficiently long evolution time and the doped final state can be efficiently prepared.  
For stronger potential, the overlap remains nearly zero and the adiabatic doping proposal fails for free fermions.  
}  
\label{fig:1d-particledope-overlap-free}
\end{figure}

We first study the adiabatic particle doping of free fermions. 
This corresponds to setting the filling factor $f$ larger than $1/2$. 
We simulate the adiabatic evolution with rational filling factors $f = 2/3, 3/4$ and calculate the wave function overlap of the final state with the ground state (Eq.~\eqref{eq:overlap}). 
The dependence of the wave function overlap on the total evolution time $T$ is shown in Fig.~\ref{fig:1d-particledope-overlap-free}. 
The initial state is a one-dimensional band insulator without interaction in optical lattice with $L$ periods.   
We consider two different choices of system size, $L = 36$, and $60$.
For current and following calculations, we use periodical boundary condition for non-interacting fermions, while the open boundary condition is adopted for the DMRG simulation of interacting fermions for numerical implementation convenience. 
For weak lattice potential, say $V = 1, 2 E_R$($E_R$ is the single photon recoil energy), the overlap increases with $T$ and quickly approaches to $1$ as we increase the total evolution time. 
This implies the adiabatic preparation of the final state  is efficient for the weak lattice confinement.  
However, with strong lattice potential ($V = 8E_R$, for example), the final state overlap essentially remains at  zero for all evolution time we have simulated, which means the adiabatic doping is inefficient.

\begin{figure}
\begin{center}
\includegraphics[width=1.0\linewidth]{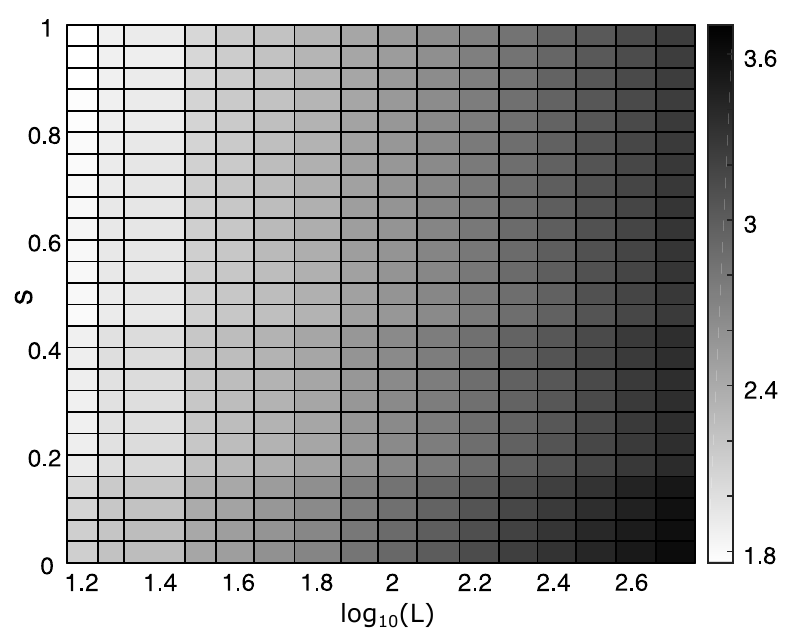} 
\end{center}
\caption{
The dependence of the quantity $\text{log}_{10} M(L)$ (Eq.~\eqref{eq:M(L)}) on the system size $L$ for the adiabatic particle doping of one-dimensional free fermions. 
Here $s\in [0,1]$ is the path parameter of adiabatic evolution (Eq.~\eqref{eq:evolutional potential}).  
The system size $L$ takes values from $15$ to $597$ with a rational filling factor $f=2/3$. 
In the calculation, the potential strength is $V=V'=8E_R$, for which the intermediate regime of evolution is strongly localized. 
It is evident that this quantity increases quickly with $L$ at each time instant $s$. 
For a one-dimensional localized system, $M(L)$ is expected to be nearly independent of $L$ due to the $L^{-1}$ scaling of $\langle {\rm NPR} \rangle$. 
The increase of $\text{log}_{10} M(L)$ implies that the localization becomes unstable as we increase the system size and tends to disappear in the thermodynamic limit.  
Nonetheless, the significant deviation of M(L) from L already implies the coupling between wave functions is rather weak, which affects the quantum adiabatic doping. 
}
\label{fig:1d-particledope-NPR-free}
\end{figure}

For the adiabatic doping with incommensurate lattice, the localization in the intermediate dynamics leads to slowing down of the adiabatic state evolution. 
This has been shown by the increase of inverse participation ratio(IPR) with the lattice potential strength~\cite{lin2019quantum}. 
In the commensurate case, similarly, the breakdown of adiabatic state preparation under strong lattice confinement indicates a localization-like problem also occurs during evolution. 
To investigate the slowing down problem in the commensurate lattice, we calculate the normalized participation ratio(NPR) \cite{li2017mobility}. 
The NPR of a single-particle eigenstate $\phi_m(x)$ is defined as
\be
\text{NPR}^{(m)} = \left[ L\sum_j |\phi_m(j)|^4 \right]^{-1},
\ee
where $L$ is the system size, and $j$ labels the discrete spatial coordinate(see Eq.~\eqref{eq:Hamiltonian-discrete}).   
This quantity remains finite for spatial extended states but vanishes for localized states. 
For a one-dimensional localized system with length $L$, it goes like $L^{-1}$. 
We calculate the single-particle NPR of $L$ lowest eigenstates and average them to get 
\be
\langle \text{NPR} \rangle = \frac{1}{L}\sum_{m= 1}^{L} \text{NPR}^{(m)}.
\ee
The averaged NPR is multiplied by $L$ to compensate the $L^{-1}$ scaling:  
\be
M(L) =  L\langle \text{NPR} \rangle.
\label{eq:M(L)}
\ee
Therefore, for localized one-dimensional system, $M(L)$ is expected to be nearly independent of $L$.  
In Fig.~\ref{fig:1d-particledope-NPR-free} we show the dependence of the quantity $\text{log}_{10}M(L)$ on $L$ following the evolution path.  
The filling factor is set to be $f=3/4$ and the overall potential strength is $V = V' = 8~E_{R}$. 
It is evident that $M(L)$ increases quickly with $L$ at each point of the path. 
While this is consistent with the well-known fact that commensurate lattice models do not have localization in the thermodynamic limit, having a $\langle {\rm NPR} \rangle$ significantly smaller than 1 in a finite size system implies the coupling between different modes is severely suppressed owing to the locality of the Hamiltonian, which then causes the slowing down of the quantum adiabatic evolution.

\subsection{Interacting fermion}

\begin{figure}
\begin{center}
\includegraphics[width=0.8\linewidth]{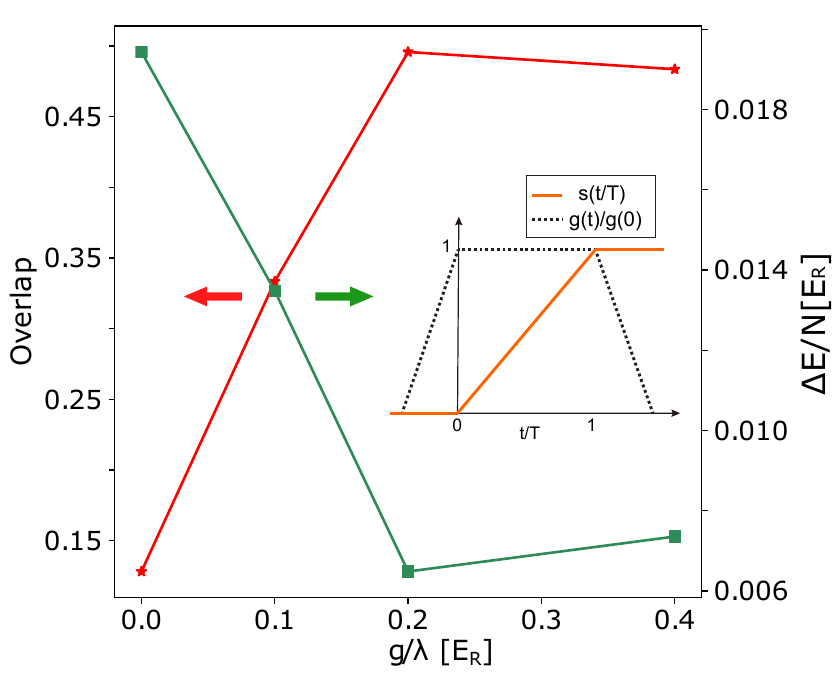}
\end{center}
\caption{
Quantum adiabatic particle doping of one-dimensional interacting fermions with rational filling. 
We simulate the adiabatic evolution with DMRG, taking the Hamiltonian in Eq.~\eqref{eq:Hamiltonian-discrete}. 
The inset shows the time sequence of the evolution. 
The interaction is adiabatically tuned on for a one-dimensional noninteracting band insulator, held constant to implement the lattice conversion, and then adiabatically tuned off. 
We perform the calculation of the lattice conversion process in the regime $t \in [0,T]$. 
In the DMRG calculation, we take second-order Trotterization of the evolution operator, with the evolution time $T = 200~\hbar/E_R$ divided into $26000$ Trotter steps. 
The filling factor is chosen as $f= 2/3 $ with $L = 14$ and $L' = 21$. 
The strength of the potential is $V = V' = 8~E_R$. 
As we increase the interaction strength from $g=0$ to $g=0.2 \lambda E_R$, the final state wave function overlap increases and the single-particle excitation energy $\Delta E/N$ ($N$ is particle number) becomes smaller.
The localization problem is reduced and the performance of the adiabatic doping is improved by introducing interaction of proper strength. }
\label{fig:par-interacting}
\end{figure}

The localization makes the minimal energy gap between the ground state and the first excited state exponentially small, and therefore causes slowing down of the adiabatic state preparation. 
In the study of many-body localization, it has been established that interaction effect tends to destabilize localization \cite{basko2006metal,oganesyan2007localization,pal_many-body_2010,kondov2015,schreiber2015observation,smith2016many,choi2016exploring}, which would then improve the efficiency of the quantum adiabatic doping. 
This has been shown to be efficient for an incommensurate lattice with strong interaction in Ref.~\onlinecite{lin2019quantum}. 
In this work, we consider introducing a time-dependent interaction 
\be 
H_{\text{U}} = g(t) \int dx \psi^{\dag}_{\uparrow}\psi^{\dag}_{\downarrow}
\psi_{\downarrow}\psi_{\uparrow}
\ee
with the time sequence shown in the inset of Fig.~\ref{fig:par-interacting}. 
The system is initially prepared in a one-dimensional noninteracting band insulator with $14$ periods. 
The interaction is turned on slowly and ramped to a constant.  
Then the initial optical lattice is adiabatically converted to another spatial period, after which the interaction is turned off slowly.  
The size of the final lattice is $L'=21$, which corresponds to a rational filling factor $f=2/3$. 
We simulate the lattice conversion process using density matrix renormalization group (DMRG) method. 
The total evolution time is $T = 200~\hbar/E_R$. We use the second-order Trotterization \cite{white2004real} of the evolution operator with $26000$ evolution steps. 
The wave function overlap and the single-particle excitation energy at $t=T$ are shown as functions of the interaction strength $g$ in Fig.~\ref{fig:par-interacting}. 
Here the overall potential strength is $V = V' = 8 E_R$, for which the non-interacting quantum adiabatic doping is inefficient.  
As we increase the interaction strength $g$ from $0$ to $0.2$ (in unit of $\lambda E_R$ with $\lambda$ the spatial period of the initial lattice), the wave function overlap is improved from $0.128$ to $0.484$, and the single-particle excitation energy $\Delta E/N$ ($N$ is the number of particles) becomes much suppressed.  
We thus find that in the commensurate case, the slowing down problem can still be solved by introducing atomic interaction. 
An dramatically enhancement of the preparation efficiency is achieved with a proper interaction strength. 
However, when we further increase the interaction strength, we find that the overlap decreases and $\Delta E/N$ increases slightly.  
In the dynamical simulation, the largest bond dimension of the matrix product state(MPS) is $\chi = 120$, with which 
numerical convergence is reached (see Ref.~\onlinecite{lin2019quantum})

\begin{figure}
\begin{center}
\includegraphics[width=1.0\linewidth]{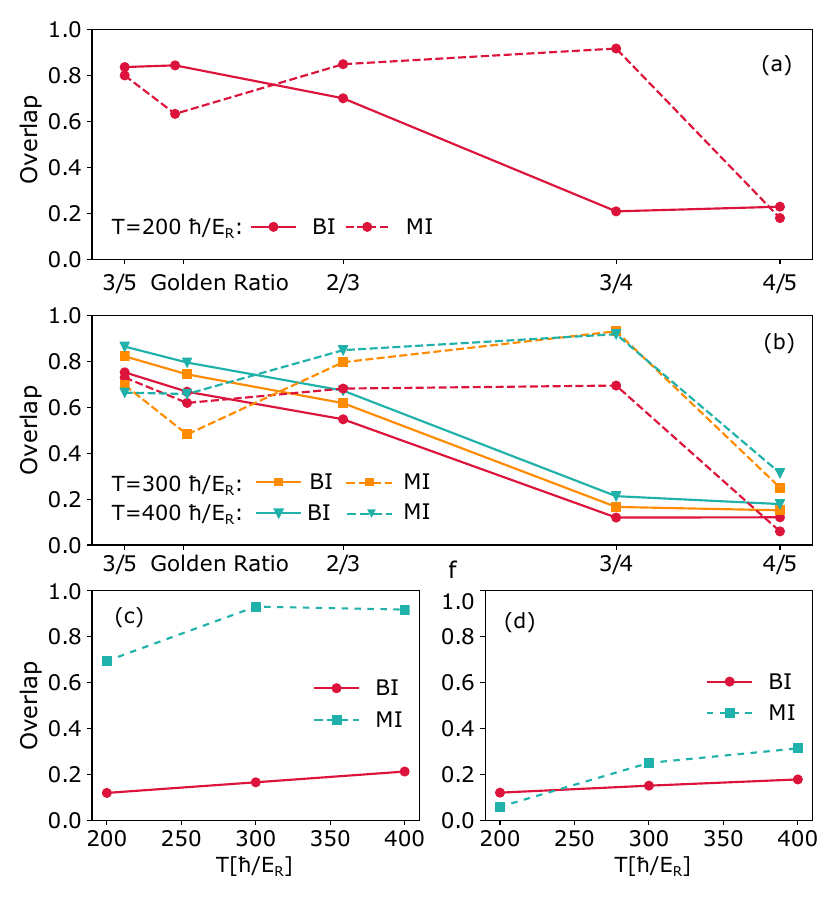}
\end{center}
\caption{
Performance of quantum adiabatic doping of one-dimensional interacting fermions starting from band insulator  and Mott insulator. 
We simulate the adiabatic evolution using DMRG with several different system sizes, $L=8,9,12,13$. 
For the Mott insulator, the corresponding number of lattice spatial periods is $2L$ with the lattice constant $\lambda/2$. 
During the lattice conversion, the interaction strength is held constant of $g=0.4\lambda E_R$. 
The final state wave function overlap of these two different protocols
are compared in (a) for $L=8,9$ with the evolution time $T=200~\hbar/E_R$ and (b) for $L=12,13$ with $T=200,300,400~\hbar/E_R$. 
We choose $5$ different final state fillings, both rational and irrational. 
In (c) and (d), we show the dependence of wave function overlap on the adiabatic time $T$ with $f=3/4$ and $4/5$, respectively. 
For $f=2/3$ and $3/4$, the adiabatic doping tarting from the Mott insulator has better performance. 
The comparison is most dramatic for $f=3/4$, as shown in (c). 
For $f=4/5$, the wave function overlaps of both procedures are smaller than $1/2$ in the time regime we consider. 
As we increase $T$, the protocol starting from a Mott insulator is more efficient than that from a band insulator.
} 
\label{fig:BI-vs-MI}
\end{figure}

We also consider performing quantum adiabatic doping starting from a Mott insulating state. 
In experiments, such an initial state is accessible since the low-temperature antiferromagnetic order has been observed at half filling in two-dimensional optical lattices \cite{mazurenko_cold-atom_2017}.
To compare the performances of starting from the two different initial states of Mott and band insulators, we simulate the adiabatic lattice conversion starting from interacting band insulator with $L$ periods and Mott insulator with $2L$ periods. 
We choose five different fillings in the particle-doped regime, both rational and irrational. 
Without loss of generality, the irrational filling factor is set to be golden ratio, which is approximated by the Fibonacci sequence as $8/13$ and $13/21$ in our finite-size calculation. 
Here we consider several different system sizes, $L=8,9,12,13$.
For the Mott insulator, each period of the initial lattice is divided into $10$ grids, while that for band insulator is divided into $20$ grids.
This parameter choice is chosen such that the final states of two procedures have the same spatial periods and discrete grids. 
The total evolution time is $T=200~\hbar/E_R$ for the smaller system of $L=8,9$ and $T = 200,300,400~\hbar/E_R$ with the corresponding numbers of Trotter steps $26000$, $38000$, and $50000$ for the larger one of $L=12,13$.  
During the evolution, the interaction strength is held constant of $g=0.4 \lambda E_R$ and the overall potential strength takes the value $V=V'=8~E_R$. 
The final state wave function overlap is shown in Fig.~\ref{fig:BI-vs-MI} with the solid and dashed lines representing the results for band insulator and Mott insulator, respectively.  
The overlap is larger for band insulator at $f=3/5$ and $(\sqrt{5}-1)/2$ . 

For $f=2/3$ and $3/4$, the adiabatic doping starting from Mott insulator has better performance. 
The advantage is especially evident for $f=3/4$. 
As shown in Fig.~\ref{fig:BI-vs-MI}(c), the overlap reaches $0.93$ at $T=300~\hbar/E_R$ for Mott insulator while that for band insulator is $0.17$.  
At $f=4/5$, the overlaps of the two protocols are both smaller than $1/2$ for all adiabatic time we have simulated.  
As shown in Fig.~\ref{fig:BI-vs-MI}(d), the overlap staring from Mott insulator exceeds that of band insulator and reaches $0.31$ at $T=400~\hbar/E_R$. 
We expect the overall performance can be further improved by increasing the total evolution time.

\section{Hole doping}

\subsection{Free fermion}

\begin{figure}
\begin{center}
\includegraphics[width=1.0\linewidth]{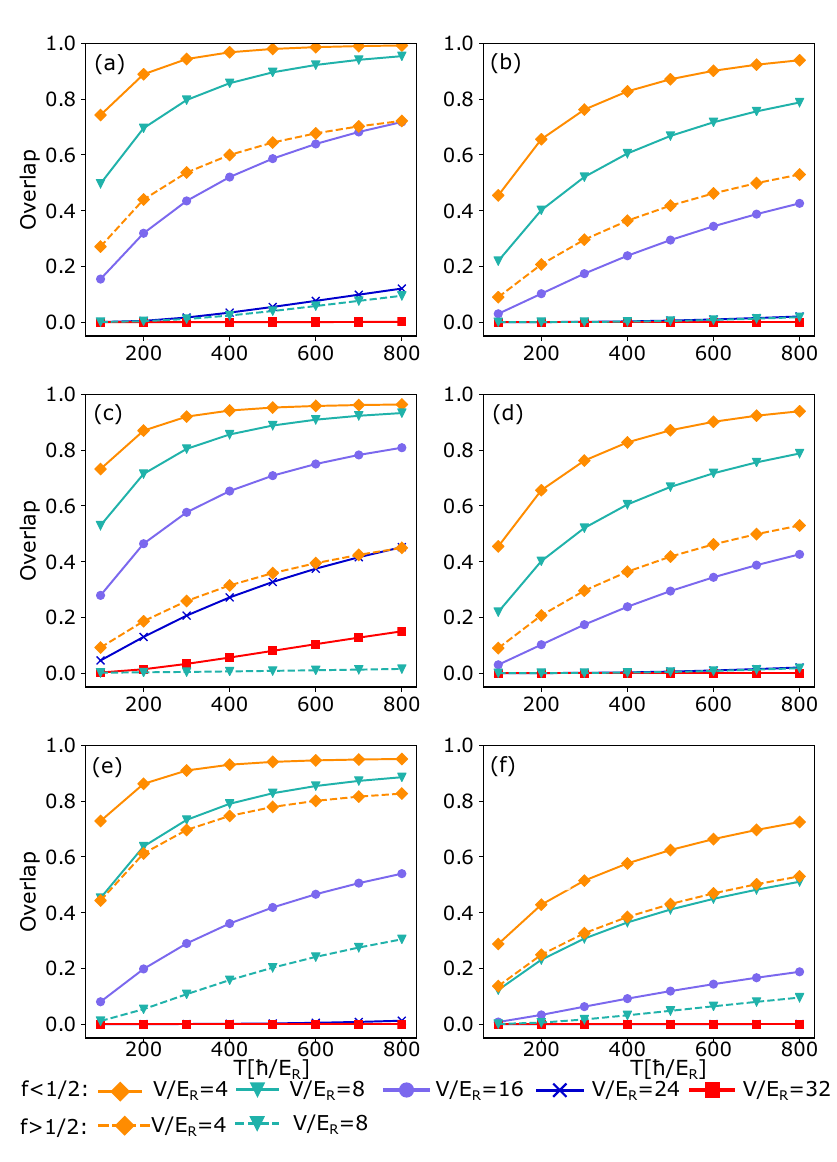}
\end{center}
\caption{ 
Quantum adiabatic hole doping of free fermions starting from one-dimensional band insulator. 
The solid and dashed lines show the dependence of final state wave function overlap on total evolution time $T$ for hole doping and particle doping, respectively. 
For hole doping, the filling factor takes rational values $1/3$ and$1/4$ in (a), (b) and (c), (d), respectively. 
The system sizes are set to be $L=36$ in (a), (c) and $L=60$ in (b), (d). 
It also takes irrational value $\left[3- \sqrt{5}\right]/2$, which is approximate by the Fibonacci sequence as $f=34/89$ with $L=34$ in (e) and $55/144$ with $L=55$ in (f). 
The non-interacting hole doping is efficient for weak lattice confinement, say $V=4E_R$. 
That is, the wave function overlap quickly approaches to $1$ for rational fillings we considered, and also the smaller system of $L=34$ with irrational filling. 
Under strong lattice confinement with $V=32E_R$, the overlap remains small and the quantum adiabatic doping fails to prepare the final state. 
For the same strength, the adiabatic doping is more efficient in the hole-doped regime.
For the fillings in hole-doped regime we consider, a overlap larger than $1/2$ can be reached at $T=800\hbar/E_R$ for $L=34,36$ with $V=16E_R$ and $L=55,60$ with $V=8E_R$, while particle doping is inefficient with $V=8E_R$ (shown by dashed lines with the corresponding filling factors $1-f$).  
}  
\label{fig:1d-holedope-overlap-free}
\end{figure}

In this section we study the adiabatic hole doping, which corresponds to the filling factor smaller than $1/2$. 
We first simulate the adiabatic evolution of free fermions starting from one-dimensional non-interacting band insulator in an optical lattice with $L$ periods. 
Both the rational and irrational fillings are considered. 
The rational filling factors are chosen as $f=1/3$ and $1/4$, while for the irrational case, we take $f=\left[3- \sqrt{5}\right]/2$, which is approximated by Fibonacci sequence as $f = F_{n-2}/F_n$. 
We choose four different values for the system size, $L=36,60$ for the rational fillings and $L=34,55$ for the irrational case. 
The dependence of the final state wave function overlap on the total evolution time is shown in Fig.~\ref{fig:1d-holedope-overlap-free}. 
The overlap systemically increases with evolution time for all potential strengths we consider. 
It should be noticed that a wave function overlap larger than $1/2$ can be reached for the larger system of $L=55,60$ with $V=8E_R$ and smaller system of $L=34,36$ with $V =16E_R$ as we increase the evolution time to $T = 800 \hbar/E_R$. 
The adiabatic doping remains efficient in the hole-doped regime even for moderate lattice potentials, for example with $V = 8E_R, 16 E_R$, in contrast to the particle-doping, which is severely subjected to localization slowing down problem. 
For sufficiently strong lattice potential, the slowing down problem still occurs in the quantum adiabatic evolution of the hole-doped case.

\begin{figure}
\begin{center}
\includegraphics[width=1.0\linewidth]{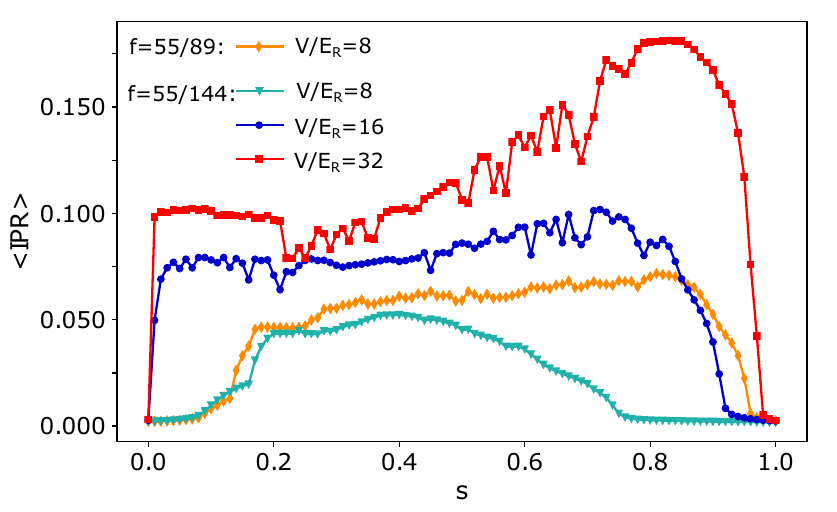}
\end{center}
\caption{ 
Averaged inverse participation ratio(IPR) along the evolution path with $V=8,16,32E_R$ in the hole-doped regime and $V=8E_R$ in the particle-doped regime. 
The filling factor is set to be $f=\left[3- \sqrt{5}\right]/2$ and $\left[\sqrt{5}-1\right]/2$ and approximated by Fibonacci sequence as $F_{n-2}/F_n$ and $F_{n-2}/F_{n-1}$ with the initial lattice size $L=55$, respectively. 
The averaged IPR increases with potential strength, which implies the localization becomes stronger. 
Compared to particle doping, the IPR for hole doping is smaller for most part of the evolution path, which explains the high preparation efficiency in this regime. 
} 
\label{fig:IPR-particlevshole}
\end{figure}

To investigate the localization in the hole-doped regime, we calculate the inverse participation ratio(IPR), which tends to vanish in the extended system and remains finite in the localized system. 
The IPR is averaged over the $L$ lowest lying single-particle eigenstates following the evolution path.
The results are shown in Fig.~\ref{fig:IPR-particlevshole}.
The irrational filling is set to be $f=\left[3- \sqrt{5}\right]/2$, which is approximated by $55/144$ with $L=55$.  
The averaged IPR systematically increases with the potential strength, 
which means the breakdown of adiabatic preparation corresponds to the atom localization. 
The results are compared to that in the particle-doped regime with $f = \left[\sqrt{5}-1\right]/2$ (approximated by $55/89$) and $V = 8E_R$. 
For the most part of Hamiltonian evolution path, the IPR of hole doping is smaller than that of particle doping, especially for $s>0.5$ where the strength of final potential becomes dominant. 
This is because in comparison with the particle-doping case, the final lattice constant is smaller for hole doping, for which quantum tunneling is stronger, suppressing the atom localization.  
The quantum adiabatic doping protocol is evidently more efficient in the case of hole-doping than the particle doping.

\subsection{Interacting fermion} 

\begin{figure}
\begin{center}
\includegraphics[width=1.0\linewidth]{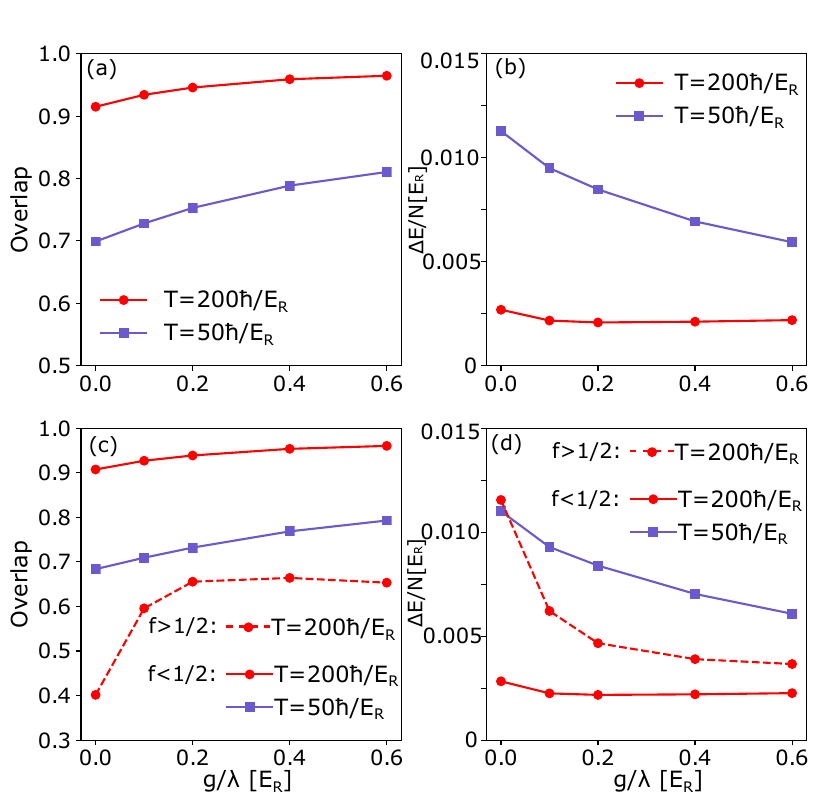}
\end{center}
\caption{ 
Performance of adiabatic hole doping of interacting fermions in one-dimensional lattice for both rational ($f = 2/5$) and irrational fillings($\left[3- \sqrt{5}\right]/2$). 
We simulate the dynamics using DMRG with the same parameter choice as in Fig~\ref{fig:par-interacting}. 
The wave function overlap and single particle excitation energy are shown 
with solid lines in (a),(b) for rational filling with $L=12$ and (c),(d) for irrational filling, which is approximated by $L/L'=13/34$.  
Here we consider the total adiabatic time $T=50\hbar/E_R$ and $200\hbar/E_R$. 
The overall potential strength is $V=8E_R$ for both particle and hole doping. 
It is found that the preparation efficiency is improved by introducing strong interaction.
As we increase the interaction strength, the wave function increases and approaches to $1$ for $T=200\hbar/E_R$, and the excitation energy is reduced accordingly.   
Compared to particle doping (shown by dashed lines with filling factor $f=13/21$),  
the hole doping shows a better preparation efficiency for this interacting case, as is true for free fermions as well.    
} 
\label{fig:holedop-overlap-interacting}
\end{figure}

We further consider adiabatic hole doping of interacting fermions, and simulate the evolution process starting from a one-dimensional interacting band insulator using DMRG. 
We choose $f=2/5$ with $L=12$ for the rational filling and $f=\left[3- \sqrt{5}\right]/2$ for the irrational one, which is approximated by $L/L'=13/34$ in the calculation.  
The dependence of the final state wave function overlap and single particle excitation energy on the interaction strength is shown in Fig.~\ref{fig:holedop-overlap-interacting} (solid lines) with an overall potential strength $V=8E_R$.  
The DMRG calculation shares the same time sequence and parameter choice with that in the particle-doped regime.
Two choices of total evolution time, $T=50\hbar/E_R$ and $200\hbar/E_R$, are considered with $7000$ and $26000$ evolution steps, respectively. 
The preparation efficiency is improved by introducing strong atomic interaction. 
In our numerical results, we find systematic increase of wave function overlap (Eq.~\eqref{eq:overlap}) and reduction of excitation energy with increasing interaction strength.
By increasing interaction strength $g$ from $0$ to $0.6$ (in unit of $\lambda E_R$), the wave function overlap  ultimately reaches $0.96$ for $f=13/34$ and $0.965$ for $f=2/5$. 
We compare the results with that of particle doping (dashed line) for the same interaction and confinement potential strengths. 
As shown in Fig.~\ref{fig:holedop-overlap-interacting} (c), the wave function overlap for $T=50\hbar/E_R$ in the hole-doped regime is even large than that of particle doping for $T=200\hbar/E_R$. 
The hole doping is evidently more efficient than the particle doping in the interacting regime, as is true for non-interacting case as well.

\medskip 
\section{Conclusion}
To conclude, the doped quantum phase of atomic Fermi-Hubbard model with low thermal entropy can be prepared by adiabatically converting two optical lattices with different spatial periods. 
In this adiabatic doping proposal, an arbitrary filling fraction can be achieved by choosing the lattice constant ratio of the initial and final lattices. 
In this work, we consider the quantum adiabatic doping starting from band insulator in one-dimensional lattice and systemically study the proposal for a broad range of filing fractions from particle-doping to hole-doping, including both rational and irrational cases. 
It is found that the atom localization, which is a fundamental problem in incommensurate lattice, also  prevents efficient adiabatic doping for commensurate filling at strong lattice confinement.
Through DMRG simulation, we show that the localization is suppressed by introducing strong atomic interaction and the state preparation efficiency 
is consequently improved.
Compared to particle doping, the adiabatic hole doping is more efficient in both free and interacting regime.
We also consider the adiabatic doping starting from the Mott insulator state at half filling.  
From the results of DMRG calculation, we find that this protocol has significantly higher preparation efficiency than the adiabatic doping starting from band insulator for a certain range of filling factors. 

Although the numerical simulation of the quantum adiabatic doping process is restricted to one dimension in this work due to numerical cost, 
we anticipate a better performance in two dimensions for the localization physics is largely expected to be weaker in higher dimensions. Our systematic study of different fillings and interaction effects should also shed light on the quantum adiabatic doping in two dimensions.

\medskip 
{\bf Acknowledgement.}
This work is supported by National Key R\&D Program of China (under grant no. 2018YFA0306502), National Natural Science Foundation of China (Grants No. 11934002 and 11774067), National Program on Key Basic Research Project of China (Grant No. 2017YFA0304204), and Shanghai Municipal Science and Technology Major Project (Grant No. 2019SHZDZX01).    
Calculations were performed based on the ITensor Library~\cite{ITensor}.

\bibliography{reference}

\begin{thebibliography}{32}%
\makeatletter
\providecommand \@ifxundefined [1]{%
 \@ifx{#1\undefined}
}%
\providecommand \@ifnum [1]{%
 \ifnum #1\expandafter \@firstoftwo
 \else \expandafter \@secondoftwo
 \fi
}%
\providecommand \@ifx [1]{%
 \ifx #1\expandafter \@firstoftwo
 \else \expandafter \@secondoftwo
 \fi
}%
\providecommand \natexlab [1]{#1}%
\providecommand \enquote  [1]{``#1''}%
\providecommand \bibnamefont  [1]{#1}%
\providecommand \bibfnamefont [1]{#1}%
\providecommand \citenamefont [1]{#1}%
\providecommand \href@noop [0]{\@secondoftwo}%
\providecommand \href [0]{\begingroup \@sanitize@url \@href}%
\providecommand \@href[1]{\@@startlink{#1}\@@href}%
\providecommand \@@href[1]{\endgroup#1\@@endlink}%
\providecommand \@sanitize@url [0]{\catcode `\\12\catcode `\$12\catcode
  `\&12\catcode `\#12\catcode `\^12\catcode `\_12\catcode `\%12\relax}%
\providecommand \@@startlink[1]{}%
\providecommand \@@endlink[0]{}%
\providecommand \url  [0]{\begingroup\@sanitize@url \@url }%
\providecommand \@url [1]{\endgroup\@href {#1}{\urlprefix }}%
\providecommand \urlprefix  [0]{URL }%
\providecommand \Eprint [0]{\href }%
\providecommand \doibase [0]{http://dx.doi.org/}%
\providecommand \selectlanguage [0]{\@gobble}%
\providecommand \bibinfo  [0]{\@secondoftwo}%
\providecommand \bibfield  [0]{\@secondoftwo}%
\providecommand \translation [1]{[#1]}%
\providecommand \BibitemOpen [0]{}%
\providecommand \bibitemStop [0]{}%
\providecommand \bibitemNoStop [0]{.\EOS\space}%
\providecommand \EOS [0]{\spacefactor3000\relax}%
\providecommand \BibitemShut  [1]{\csname bibitem#1\endcsname}%
\let\auto@bib@innerbib\@empty
\bibitem [{\citenamefont {Lin}\ \emph {et~al.}(2019)\citenamefont {Lin},
  \citenamefont {Nan}, \citenamefont {Luo}, \citenamefont {Yao},\ and\
  \citenamefont {Li}}]{lin2019quantum}%
  \BibitemOpen
  \bibfield  {author} {\bibinfo {author} {\bibfnamefont {J.}~\bibnamefont
  {Lin}}, \bibinfo {author} {\bibfnamefont {J.}~\bibnamefont {Nan}}, \bibinfo
  {author} {\bibfnamefont {Y.}~\bibnamefont {Luo}}, \bibinfo {author}
  {\bibfnamefont {X.-C.}\ \bibnamefont {Yao}}, \ and\ \bibinfo {author}
  {\bibfnamefont {X.}~\bibnamefont {Li}},\ }\href@noop {} {\bibfield  {journal}
  {\bibinfo  {journal} {Physical Review Letters}\ }\textbf {\bibinfo {volume}
  {123}},\ \bibinfo {pages} {233603} (\bibinfo {year} {2019})}\BibitemShut
  {NoStop}%
\bibitem [{\citenamefont {Bloch}\ \emph {et~al.}(2008)\citenamefont {Bloch},
  \citenamefont {Dalibard},\ and\ \citenamefont {Zwerger}}]{bloch2008many}%
  \BibitemOpen
  \bibfield  {author} {\bibinfo {author} {\bibfnamefont {I.}~\bibnamefont
  {Bloch}}, \bibinfo {author} {\bibfnamefont {J.}~\bibnamefont {Dalibard}}, \
  and\ \bibinfo {author} {\bibfnamefont {W.}~\bibnamefont {Zwerger}},\
  }\href@noop {} {\bibfield  {journal} {\bibinfo  {journal} {Reviews of modern
  physics}\ }\textbf {\bibinfo {volume} {80}},\ \bibinfo {pages} {885}
  (\bibinfo {year} {2008})}\BibitemShut {NoStop}%
\bibitem [{\citenamefont {Li}\ and\ \citenamefont {Liu}(2016)}]{li2016physics}%
  \BibitemOpen
  \bibfield  {author} {\bibinfo {author} {\bibfnamefont {X.}~\bibnamefont
  {Li}}\ and\ \bibinfo {author} {\bibfnamefont {W.~V.}\ \bibnamefont {Liu}},\
  }\href@noop {} {\bibfield  {journal} {\bibinfo  {journal} {Reports on
  Progress in Physics}\ }\textbf {\bibinfo {volume} {79}},\ \bibinfo {pages}
  {116401} (\bibinfo {year} {2016})}\BibitemShut {NoStop}%
\bibitem [{\citenamefont {Lewenstein}\ \emph {et~al.}(2007)\citenamefont
  {Lewenstein}, \citenamefont {Sanpera}, \citenamefont {Ahufinger},
  \citenamefont {Damski}, \citenamefont {Sen},\ and\ \citenamefont
  {Sen}}]{lewenstein2007ultracold}%
  \BibitemOpen
  \bibfield  {author} {\bibinfo {author} {\bibfnamefont {M.}~\bibnamefont
  {Lewenstein}}, \bibinfo {author} {\bibfnamefont {A.}~\bibnamefont {Sanpera}},
  \bibinfo {author} {\bibfnamefont {V.}~\bibnamefont {Ahufinger}}, \bibinfo
  {author} {\bibfnamefont {B.}~\bibnamefont {Damski}}, \bibinfo {author}
  {\bibfnamefont {A.}~\bibnamefont {Sen}}, \ and\ \bibinfo {author}
  {\bibfnamefont {U.}~\bibnamefont {Sen}},\ }\href@noop {} {\bibfield
  {journal} {\bibinfo  {journal} {Advances in Physics}\ }\textbf {\bibinfo
  {volume} {56}},\ \bibinfo {pages} {243} (\bibinfo {year} {2007})}\BibitemShut
  {NoStop}%
\bibitem [{\citenamefont {Bloch}(2018)}]{bloch2018quantum}%
  \BibitemOpen
  \bibfield  {author} {\bibinfo {author} {\bibfnamefont {I.}~\bibnamefont
  {Bloch}},\ }\href@noop {} {\bibfield  {journal} {\bibinfo  {journal} {Nature
  Physics}\ }\textbf {\bibinfo {volume} {14}},\ \bibinfo {pages} {1159}
  (\bibinfo {year} {2018})}\BibitemShut {NoStop}%
\bibitem [{\citenamefont {Jaksch}\ \emph {et~al.}(1998)\citenamefont {Jaksch},
  \citenamefont {Bruder}, \citenamefont {Cirac}, \citenamefont {Gardiner},\
  and\ \citenamefont {Zoller}}]{jaksch1998}%
  \BibitemOpen
  \bibfield  {author} {\bibinfo {author} {\bibfnamefont {D.}~\bibnamefont
  {Jaksch}}, \bibinfo {author} {\bibfnamefont {C.}~\bibnamefont {Bruder}},
  \bibinfo {author} {\bibfnamefont {J.~I.}\ \bibnamefont {Cirac}}, \bibinfo
  {author} {\bibfnamefont {C.~W.}\ \bibnamefont {Gardiner}}, \ and\ \bibinfo
  {author} {\bibfnamefont {P.}~\bibnamefont {Zoller}},\ }\href@noop {}
  {\bibfield  {journal} {\bibinfo  {journal} {Phys. Rev. Lett.}\ }\textbf
  {\bibinfo {volume} {81}},\ \bibinfo {pages} {3108} (\bibinfo {year}
  {1998})}\BibitemShut {NoStop}%
\bibitem [{\citenamefont {Hofstetter}\ \emph {et~al.}(2002)\citenamefont
  {Hofstetter}, \citenamefont {Cirac}, \citenamefont {Zoller}, \citenamefont
  {Demler},\ and\ \citenamefont {Lukin}}]{hofstetter2002high}%
  \BibitemOpen
  \bibfield  {author} {\bibinfo {author} {\bibfnamefont {W.}~\bibnamefont
  {Hofstetter}}, \bibinfo {author} {\bibfnamefont {J.~I.}\ \bibnamefont
  {Cirac}}, \bibinfo {author} {\bibfnamefont {P.}~\bibnamefont {Zoller}},
  \bibinfo {author} {\bibfnamefont {E.}~\bibnamefont {Demler}}, \ and\ \bibinfo
  {author} {\bibfnamefont {M.}~\bibnamefont {Lukin}},\ }\href@noop {}
  {\bibfield  {journal} {\bibinfo  {journal} {Physical review letters}\
  }\textbf {\bibinfo {volume} {89}},\ \bibinfo {pages} {220407} (\bibinfo
  {year} {2002})}\BibitemShut {NoStop}%
\bibitem [{\citenamefont {Dutta}\ \emph {et~al.}(2015)\citenamefont {Dutta},
  \citenamefont {Gajda}, \citenamefont {Hauke}, \citenamefont {Lewenstein},
  \citenamefont {L{\"u}hmann}, \citenamefont {Malomed}, \citenamefont
  {Sowi{\'n}ski},\ and\ \citenamefont {Zakrzewski}}]{dutta2015non}%
  \BibitemOpen
  \bibfield  {author} {\bibinfo {author} {\bibfnamefont {O.}~\bibnamefont
  {Dutta}}, \bibinfo {author} {\bibfnamefont {M.}~\bibnamefont {Gajda}},
  \bibinfo {author} {\bibfnamefont {P.}~\bibnamefont {Hauke}}, \bibinfo
  {author} {\bibfnamefont {M.}~\bibnamefont {Lewenstein}}, \bibinfo {author}
  {\bibfnamefont {D.-S.}\ \bibnamefont {L{\"u}hmann}}, \bibinfo {author}
  {\bibfnamefont {B.~A.}\ \bibnamefont {Malomed}}, \bibinfo {author}
  {\bibfnamefont {T.}~\bibnamefont {Sowi{\'n}ski}}, \ and\ \bibinfo {author}
  {\bibfnamefont {J.}~\bibnamefont {Zakrzewski}},\ }\href@noop {} {\bibfield
  {journal} {\bibinfo  {journal} {Reports on Progress in Physics}\ }\textbf
  {\bibinfo {volume} {78}},\ \bibinfo {pages} {066001} (\bibinfo {year}
  {2015})}\BibitemShut {NoStop}%
\bibitem [{\citenamefont {Lee}\ \emph {et~al.}(2006)\citenamefont {Lee},
  \citenamefont {Nagaosa},\ and\ \citenamefont {Wen}}]{lee2006doping}%
  \BibitemOpen
  \bibfield  {author} {\bibinfo {author} {\bibfnamefont {P.~A.}\ \bibnamefont
  {Lee}}, \bibinfo {author} {\bibfnamefont {N.}~\bibnamefont {Nagaosa}}, \ and\
  \bibinfo {author} {\bibfnamefont {X.-G.}\ \bibnamefont {Wen}},\ }\href@noop
  {} {\bibfield  {journal} {\bibinfo  {journal} {Reviews of modern physics}\
  }\textbf {\bibinfo {volume} {78}},\ \bibinfo {pages} {17} (\bibinfo {year}
  {2006})}\BibitemShut {NoStop}%
\bibitem [{\citenamefont {Metzner}\ \emph {et~al.}(2012)\citenamefont
  {Metzner}, \citenamefont {Salmhofer}, \citenamefont {Honerkamp},
  \citenamefont {Meden},\ and\ \citenamefont
  {Sch{\"o}nhammer}}]{2012_Metzner_RMP}%
  \BibitemOpen
  \bibfield  {author} {\bibinfo {author} {\bibfnamefont {W.}~\bibnamefont
  {Metzner}}, \bibinfo {author} {\bibfnamefont {M.}~\bibnamefont {Salmhofer}},
  \bibinfo {author} {\bibfnamefont {C.}~\bibnamefont {Honerkamp}}, \bibinfo
  {author} {\bibfnamefont {V.}~\bibnamefont {Meden}}, \ and\ \bibinfo {author}
  {\bibfnamefont {K.}~\bibnamefont {Sch{\"o}nhammer}},\ }\href@noop {}
  {\bibfield  {journal} {\bibinfo  {journal} {Reviews of Modern Physics}\
  }\textbf {\bibinfo {volume} {84}},\ \bibinfo {pages} {299} (\bibinfo {year}
  {2012})}\BibitemShut {NoStop}%
\bibitem [{\citenamefont {Jiang}\ and\ \citenamefont
  {Devereaux}(2019)}]{2019_Jiang_Science}%
  \BibitemOpen
  \bibfield  {author} {\bibinfo {author} {\bibfnamefont {H.-C.}\ \bibnamefont
  {Jiang}}\ and\ \bibinfo {author} {\bibfnamefont {T.~P.}\ \bibnamefont
  {Devereaux}},\ }\href@noop {} {\bibfield  {journal} {\bibinfo  {journal}
  {Science}\ }\textbf {\bibinfo {volume} {365}},\ \bibinfo {pages} {1424}
  (\bibinfo {year} {2019})}\BibitemShut {NoStop}%
\bibitem [{\citenamefont {Qin}\ \emph {et~al.}(2020)\citenamefont {Qin},
  \citenamefont {Chung}, \citenamefont {Shi}, \citenamefont {Vitali},
  \citenamefont {Hubig}, \citenamefont {Schollw\"ock}, \citenamefont {White},\
  and\ \citenamefont {Zhang}}]{2020_Qin_PRX}%
  \BibitemOpen
  \bibfield  {author} {\bibinfo {author} {\bibfnamefont {M.}~\bibnamefont
  {Qin}}, \bibinfo {author} {\bibfnamefont {C.-M.}\ \bibnamefont {Chung}},
  \bibinfo {author} {\bibfnamefont {H.}~\bibnamefont {Shi}}, \bibinfo {author}
  {\bibfnamefont {E.}~\bibnamefont {Vitali}}, \bibinfo {author} {\bibfnamefont
  {C.}~\bibnamefont {Hubig}}, \bibinfo {author} {\bibfnamefont
  {U.}~\bibnamefont {Schollw\"ock}}, \bibinfo {author} {\bibfnamefont {S.~R.}\
  \bibnamefont {White}}, \ and\ \bibinfo {author} {\bibfnamefont
  {S.}~\bibnamefont {Zhang}} (\bibinfo {collaboration} {Simons Collaboration on
  the Many-Electron Problem}),\ }\href@noop {} {\bibfield  {journal} {\bibinfo
  {journal} {Phys. Rev. X}\ }\textbf {\bibinfo {volume} {10}},\ \bibinfo
  {pages} {031016} (\bibinfo {year} {2020})}\BibitemShut {NoStop}%
\bibitem [{\citenamefont {Greif}\ \emph {et~al.}(2013)\citenamefont {Greif},
  \citenamefont {Uehlinger}, \citenamefont {Jotzu}, \citenamefont {Tarruell},\
  and\ \citenamefont {Esslinger}}]{greif2013short}%
  \BibitemOpen
  \bibfield  {author} {\bibinfo {author} {\bibfnamefont {D.}~\bibnamefont
  {Greif}}, \bibinfo {author} {\bibfnamefont {T.}~\bibnamefont {Uehlinger}},
  \bibinfo {author} {\bibfnamefont {G.}~\bibnamefont {Jotzu}}, \bibinfo
  {author} {\bibfnamefont {L.}~\bibnamefont {Tarruell}}, \ and\ \bibinfo
  {author} {\bibfnamefont {T.}~\bibnamefont {Esslinger}},\ }\href@noop {}
  {\bibfield  {journal} {\bibinfo  {journal} {Science}\ }\textbf {\bibinfo
  {volume} {340}},\ \bibinfo {pages} {1307} (\bibinfo {year}
  {2013})}\BibitemShut {NoStop}%
\bibitem [{\citenamefont {Hart}\ \emph {et~al.}(2015)\citenamefont {Hart},
  \citenamefont {Duarte}, \citenamefont {Yang}, \citenamefont {Liu},
  \citenamefont {Paiva}, \citenamefont {Khatami}, \citenamefont {Scalettar},
  \citenamefont {Trivedi}, \citenamefont {Huse},\ and\ \citenamefont
  {Hulet}}]{hart_observation_2015}%
  \BibitemOpen
  \bibfield  {author} {\bibinfo {author} {\bibfnamefont {R.~A.}\ \bibnamefont
  {Hart}}, \bibinfo {author} {\bibfnamefont {P.~M.}\ \bibnamefont {Duarte}},
  \bibinfo {author} {\bibfnamefont {T.-L.}\ \bibnamefont {Yang}}, \bibinfo
  {author} {\bibfnamefont {X.}~\bibnamefont {Liu}}, \bibinfo {author}
  {\bibfnamefont {T.}~\bibnamefont {Paiva}}, \bibinfo {author} {\bibfnamefont
  {E.}~\bibnamefont {Khatami}}, \bibinfo {author} {\bibfnamefont {R.~T.}\
  \bibnamefont {Scalettar}}, \bibinfo {author} {\bibfnamefont {N.}~\bibnamefont
  {Trivedi}}, \bibinfo {author} {\bibfnamefont {D.~A.}\ \bibnamefont {Huse}}, \
  and\ \bibinfo {author} {\bibfnamefont {R.~G.}\ \bibnamefont {Hulet}},\
  }\href@noop {} {\bibfield  {journal} {\bibinfo  {journal} {Nature}\ }\textbf
  {\bibinfo {volume} {519}},\ \bibinfo {pages} {211} (\bibinfo {year}
  {2015})}\BibitemShut {NoStop}%
\bibitem [{\citenamefont {Mazurenko}\ \emph {et~al.}(2017)\citenamefont
  {Mazurenko}, \citenamefont {Chiu}, \citenamefont {Ji}, \citenamefont
  {Parsons}, \citenamefont {Kanász-Nagy}, \citenamefont {Schmidt},
  \citenamefont {Grusdt}, \citenamefont {Demler}, \citenamefont {Greif},\ and\
  \citenamefont {Greiner}}]{mazurenko_cold-atom_2017}%
  \BibitemOpen
  \bibfield  {author} {\bibinfo {author} {\bibfnamefont {A.}~\bibnamefont
  {Mazurenko}}, \bibinfo {author} {\bibfnamefont {C.~S.}\ \bibnamefont {Chiu}},
  \bibinfo {author} {\bibfnamefont {G.}~\bibnamefont {Ji}}, \bibinfo {author}
  {\bibfnamefont {M.~F.}\ \bibnamefont {Parsons}}, \bibinfo {author}
  {\bibfnamefont {M.}~\bibnamefont {Kanász-Nagy}}, \bibinfo {author}
  {\bibfnamefont {R.}~\bibnamefont {Schmidt}}, \bibinfo {author} {\bibfnamefont
  {F.}~\bibnamefont {Grusdt}}, \bibinfo {author} {\bibfnamefont
  {E.}~\bibnamefont {Demler}}, \bibinfo {author} {\bibfnamefont
  {D.}~\bibnamefont {Greif}}, \ and\ \bibinfo {author} {\bibfnamefont
  {M.}~\bibnamefont {Greiner}},\ }\href@noop {} {\bibfield  {journal} {\bibinfo
   {journal} {Nature}\ }\textbf {\bibinfo {volume} {545}},\ \bibinfo {pages}
  {462} (\bibinfo {year} {2017})}\BibitemShut {NoStop}%
\bibitem [{\citenamefont {Trebst}\ \emph {et~al.}(2006)\citenamefont {Trebst},
  \citenamefont {Schollw\"ock}, \citenamefont {Troyer},\ and\ \citenamefont
  {Zoller}}]{zoller2006}%
  \BibitemOpen
  \bibfield  {author} {\bibinfo {author} {\bibfnamefont {S.}~\bibnamefont
  {Trebst}}, \bibinfo {author} {\bibfnamefont {U.}~\bibnamefont
  {Schollw\"ock}}, \bibinfo {author} {\bibfnamefont {M.}~\bibnamefont
  {Troyer}}, \ and\ \bibinfo {author} {\bibfnamefont {P.}~\bibnamefont
  {Zoller}},\ }\href@noop {} {\bibfield  {journal} {\bibinfo  {journal} {Phys.
  Rev. Lett.}\ }\textbf {\bibinfo {volume} {96}},\ \bibinfo {pages} {250402}
  (\bibinfo {year} {2006})}\BibitemShut {NoStop}%
\bibitem [{\citenamefont {S{\o}rensen}\ \emph {et~al.}(2010)\citenamefont
  {S{\o}rensen}, \citenamefont {Altman}, \citenamefont {Gullans}, \citenamefont
  {Porto}, \citenamefont {Lukin},\ and\ \citenamefont
  {Demler}}]{sorensen2010adiabatic}%
  \BibitemOpen
  \bibfield  {author} {\bibinfo {author} {\bibfnamefont {A.~S.}\ \bibnamefont
  {S{\o}rensen}}, \bibinfo {author} {\bibfnamefont {E.}~\bibnamefont {Altman}},
  \bibinfo {author} {\bibfnamefont {M.}~\bibnamefont {Gullans}}, \bibinfo
  {author} {\bibfnamefont {J.}~\bibnamefont {Porto}}, \bibinfo {author}
  {\bibfnamefont {M.~D.}\ \bibnamefont {Lukin}}, \ and\ \bibinfo {author}
  {\bibfnamefont {E.}~\bibnamefont {Demler}},\ }\href@noop {} {\bibfield
  {journal} {\bibinfo  {journal} {Physical Review A}\ }\textbf {\bibinfo
  {volume} {81}},\ \bibinfo {pages} {061603} (\bibinfo {year}
  {2010})}\BibitemShut {NoStop}%
\bibitem [{\citenamefont {Lubasch}\ \emph {et~al.}(2011)\citenamefont
  {Lubasch}, \citenamefont {Murg}, \citenamefont {Schneider}, \citenamefont
  {Cirac},\ and\ \citenamefont {Banuls}}]{lubasch2011adiabatic}%
  \BibitemOpen
  \bibfield  {author} {\bibinfo {author} {\bibfnamefont {M.}~\bibnamefont
  {Lubasch}}, \bibinfo {author} {\bibfnamefont {V.}~\bibnamefont {Murg}},
  \bibinfo {author} {\bibfnamefont {U.}~\bibnamefont {Schneider}}, \bibinfo
  {author} {\bibfnamefont {J.~I.}\ \bibnamefont {Cirac}}, \ and\ \bibinfo
  {author} {\bibfnamefont {M.-C.}\ \bibnamefont {Banuls}},\ }\href@noop {}
  {\bibfield  {journal} {\bibinfo  {journal} {Physical review letters}\
  }\textbf {\bibinfo {volume} {107}},\ \bibinfo {pages} {165301} (\bibinfo
  {year} {2011})}\BibitemShut {NoStop}%
\bibitem [{\citenamefont {Zhang}\ and\ \citenamefont {Duan}(2013)}]{Duan2013}%
  \BibitemOpen
  \bibfield  {author} {\bibinfo {author} {\bibfnamefont {Z.}~\bibnamefont
  {Zhang}}\ and\ \bibinfo {author} {\bibfnamefont {L.-M.}\ \bibnamefont
  {Duan}},\ }\href@noop {} {\bibfield  {journal} {\bibinfo  {journal} {Phys.
  Rev. Lett.}\ }\textbf {\bibinfo {volume} {111}},\ \bibinfo {pages} {180401}
  (\bibinfo {year} {2013})}\BibitemShut {NoStop}%
\bibitem [{\citenamefont {Chiu}\ \emph {et~al.}(2018)\citenamefont {Chiu},
  \citenamefont {Ji}, \citenamefont {Mazurenko}, \citenamefont {Greif},\ and\
  \citenamefont {Greiner}}]{Chiu2008}%
  \BibitemOpen
  \bibfield  {author} {\bibinfo {author} {\bibfnamefont {C.~S.}\ \bibnamefont
  {Chiu}}, \bibinfo {author} {\bibfnamefont {G.}~\bibnamefont {Ji}}, \bibinfo
  {author} {\bibfnamefont {A.}~\bibnamefont {Mazurenko}}, \bibinfo {author}
  {\bibfnamefont {D.}~\bibnamefont {Greif}}, \ and\ \bibinfo {author}
  {\bibfnamefont {M.}~\bibnamefont {Greiner}},\ }\href@noop {} {\bibfield
  {journal} {\bibinfo  {journal} {Phys. Rev. Lett.}\ }\textbf {\bibinfo
  {volume} {120}},\ \bibinfo {pages} {243201} (\bibinfo {year}
  {2018})}\BibitemShut {NoStop}%
\bibitem [{\citenamefont {Sun}\ \emph {et~al.}(2020)\citenamefont {Sun},
  \citenamefont {Yang}, \citenamefont {Wang}, \citenamefont {Zhou},
  \citenamefont {Su}, \citenamefont {Dai}, \citenamefont {Yuan},\ and\
  \citenamefont {Pan}}]{sun2020realization}%
  \BibitemOpen
  \bibfield  {author} {\bibinfo {author} {\bibfnamefont {H.}~\bibnamefont
  {Sun}}, \bibinfo {author} {\bibfnamefont {B.}~\bibnamefont {Yang}}, \bibinfo
  {author} {\bibfnamefont {H.-Y.}\ \bibnamefont {Wang}}, \bibinfo {author}
  {\bibfnamefont {Z.-Y.}\ \bibnamefont {Zhou}}, \bibinfo {author}
  {\bibfnamefont {G.-X.}\ \bibnamefont {Su}}, \bibinfo {author} {\bibfnamefont
  {H.-N.}\ \bibnamefont {Dai}}, \bibinfo {author} {\bibfnamefont {Z.-S.}\
  \bibnamefont {Yuan}}, \ and\ \bibinfo {author} {\bibfnamefont {J.-W.}\
  \bibnamefont {Pan}},\ }\href@noop {} {\bibfield  {journal} {\bibinfo
  {journal} {arXiv preprint arXiv:2009.01426}\ } (\bibinfo {year}
  {2020})}\BibitemShut {NoStop}%
\bibitem [{\citenamefont {Basko}\ \emph {et~al.}(2006)\citenamefont {Basko},
  \citenamefont {Aleiner},\ and\ \citenamefont {Altshuler}}]{basko2006metal}%
  \BibitemOpen
  \bibfield  {author} {\bibinfo {author} {\bibfnamefont {D.~M.}\ \bibnamefont
  {Basko}}, \bibinfo {author} {\bibfnamefont {I.~L.}\ \bibnamefont {Aleiner}},
  \ and\ \bibinfo {author} {\bibfnamefont {B.~L.}\ \bibnamefont {Altshuler}},\
  }\href@noop {} {\bibfield  {journal} {\bibinfo  {journal} {Annals of
  physics}\ }\textbf {\bibinfo {volume} {321}},\ \bibinfo {pages} {1126}
  (\bibinfo {year} {2006})}\BibitemShut {NoStop}%
\bibitem [{\citenamefont {Oganesyan}\ and\ \citenamefont
  {Huse}(2007)}]{oganesyan2007localization}%
  \BibitemOpen
  \bibfield  {author} {\bibinfo {author} {\bibfnamefont {V.}~\bibnamefont
  {Oganesyan}}\ and\ \bibinfo {author} {\bibfnamefont {D.~A.}\ \bibnamefont
  {Huse}},\ }\href@noop {} {\bibfield  {journal} {\bibinfo  {journal} {Physical
  review b}\ }\textbf {\bibinfo {volume} {75}},\ \bibinfo {pages} {155111}
  (\bibinfo {year} {2007})}\BibitemShut {NoStop}%
\bibitem [{\citenamefont {Pal}\ and\ \citenamefont
  {Huse}(2010)}]{pal_many-body_2010}%
  \BibitemOpen
  \bibfield  {author} {\bibinfo {author} {\bibfnamefont {A.}~\bibnamefont
  {Pal}}\ and\ \bibinfo {author} {\bibfnamefont {D.~A.}\ \bibnamefont {Huse}},\
  }\href@noop {} {\bibfield  {journal} {\bibinfo  {journal} {Physical Review
  B}\ }\textbf {\bibinfo {volume} {82}} (\bibinfo {year} {2010})}\BibitemShut
  {NoStop}%
\bibitem [{\citenamefont {Kondov}\ \emph {et~al.}(2015)\citenamefont {Kondov},
  \citenamefont {McGehee}, \citenamefont {Xu},\ and\ \citenamefont
  {DeMarco}}]{kondov2015}%
  \BibitemOpen
  \bibfield  {author} {\bibinfo {author} {\bibfnamefont {S.~S.}\ \bibnamefont
  {Kondov}}, \bibinfo {author} {\bibfnamefont {W.~R.}\ \bibnamefont {McGehee}},
  \bibinfo {author} {\bibfnamefont {W.}~\bibnamefont {Xu}}, \ and\ \bibinfo
  {author} {\bibfnamefont {B.}~\bibnamefont {DeMarco}},\ }\href@noop {}
  {\bibfield  {journal} {\bibinfo  {journal} {Phys. Rev. Lett.}\ }\textbf
  {\bibinfo {volume} {114}},\ \bibinfo {pages} {083002} (\bibinfo {year}
  {2015})}\BibitemShut {NoStop}%
\bibitem [{\citenamefont {Albash}\ and\ \citenamefont
  {Lidar}(2018)}]{albash2018adiabatic}%
  \BibitemOpen
  \bibfield  {author} {\bibinfo {author} {\bibfnamefont {T.}~\bibnamefont
  {Albash}}\ and\ \bibinfo {author} {\bibfnamefont {D.~A.}\ \bibnamefont
  {Lidar}},\ }\href@noop {} {\bibfield  {journal} {\bibinfo  {journal} {Reviews
  of Modern Physics}\ }\textbf {\bibinfo {volume} {90}},\ \bibinfo {pages}
  {015002} (\bibinfo {year} {2018})}\BibitemShut {NoStop}%
\bibitem [{\citenamefont {Li}\ \emph {et~al.}(2017)\citenamefont {Li},
  \citenamefont {Li},\ and\ \citenamefont {Sarma}}]{li2017mobility}%
  \BibitemOpen
  \bibfield  {author} {\bibinfo {author} {\bibfnamefont {X.}~\bibnamefont
  {Li}}, \bibinfo {author} {\bibfnamefont {X.}~\bibnamefont {Li}}, \ and\
  \bibinfo {author} {\bibfnamefont {S.~D.}\ \bibnamefont {Sarma}},\ }\href@noop
  {} {\bibfield  {journal} {\bibinfo  {journal} {Physical Review B}\ }\textbf
  {\bibinfo {volume} {96}},\ \bibinfo {pages} {085119} (\bibinfo {year}
  {2017})}\BibitemShut {NoStop}%
\bibitem [{\citenamefont {Schreiber}\ \emph {et~al.}(2015)\citenamefont
  {Schreiber}, \citenamefont {Hodgman}, \citenamefont {Bordia}, \citenamefont
  {L{\"u}schen}, \citenamefont {Fischer}, \citenamefont {Vosk}, \citenamefont
  {Altman}, \citenamefont {Schneider},\ and\ \citenamefont
  {Bloch}}]{schreiber2015observation}%
  \BibitemOpen
  \bibfield  {author} {\bibinfo {author} {\bibfnamefont {M.}~\bibnamefont
  {Schreiber}}, \bibinfo {author} {\bibfnamefont {S.~S.}\ \bibnamefont
  {Hodgman}}, \bibinfo {author} {\bibfnamefont {P.}~\bibnamefont {Bordia}},
  \bibinfo {author} {\bibfnamefont {H.~P.}\ \bibnamefont {L{\"u}schen}},
  \bibinfo {author} {\bibfnamefont {M.~H.}\ \bibnamefont {Fischer}}, \bibinfo
  {author} {\bibfnamefont {R.}~\bibnamefont {Vosk}}, \bibinfo {author}
  {\bibfnamefont {E.}~\bibnamefont {Altman}}, \bibinfo {author} {\bibfnamefont
  {U.}~\bibnamefont {Schneider}}, \ and\ \bibinfo {author} {\bibfnamefont
  {I.}~\bibnamefont {Bloch}},\ }\href@noop {} {\bibfield  {journal} {\bibinfo
  {journal} {Science}\ }\textbf {\bibinfo {volume} {349}},\ \bibinfo {pages}
  {842} (\bibinfo {year} {2015})}\BibitemShut {NoStop}%
\bibitem [{\citenamefont {Smith}\ \emph {et~al.}(2016)\citenamefont {Smith},
  \citenamefont {Lee}, \citenamefont {Richerme}, \citenamefont {Neyenhuis},
  \citenamefont {Hess}, \citenamefont {Hauke}, \citenamefont {Heyl},
  \citenamefont {Huse},\ and\ \citenamefont {Monroe}}]{smith2016many}%
  \BibitemOpen
  \bibfield  {author} {\bibinfo {author} {\bibfnamefont {J.}~\bibnamefont
  {Smith}}, \bibinfo {author} {\bibfnamefont {A.}~\bibnamefont {Lee}}, \bibinfo
  {author} {\bibfnamefont {P.}~\bibnamefont {Richerme}}, \bibinfo {author}
  {\bibfnamefont {B.}~\bibnamefont {Neyenhuis}}, \bibinfo {author}
  {\bibfnamefont {P.~W.}\ \bibnamefont {Hess}}, \bibinfo {author}
  {\bibfnamefont {P.}~\bibnamefont {Hauke}}, \bibinfo {author} {\bibfnamefont
  {M.}~\bibnamefont {Heyl}}, \bibinfo {author} {\bibfnamefont {D.~A.}\
  \bibnamefont {Huse}}, \ and\ \bibinfo {author} {\bibfnamefont
  {C.}~\bibnamefont {Monroe}},\ }\href@noop {} {\bibfield  {journal} {\bibinfo
  {journal} {Nature Physics}\ }\textbf {\bibinfo {volume} {12}},\ \bibinfo
  {pages} {907} (\bibinfo {year} {2016})}\BibitemShut {NoStop}%
\bibitem [{\citenamefont {Choi}\ \emph {et~al.}(2016)\citenamefont {Choi},
  \citenamefont {Hild}, \citenamefont {Zeiher}, \citenamefont {Schau{\ss}},
  \citenamefont {Rubio-Abadal}, \citenamefont {Yefsah}, \citenamefont
  {Khemani}, \citenamefont {Huse}, \citenamefont {Bloch},\ and\ \citenamefont
  {Gross}}]{choi2016exploring}%
  \BibitemOpen
  \bibfield  {author} {\bibinfo {author} {\bibfnamefont {J.-y.}\ \bibnamefont
  {Choi}}, \bibinfo {author} {\bibfnamefont {S.}~\bibnamefont {Hild}}, \bibinfo
  {author} {\bibfnamefont {J.}~\bibnamefont {Zeiher}}, \bibinfo {author}
  {\bibfnamefont {P.}~\bibnamefont {Schau{\ss}}}, \bibinfo {author}
  {\bibfnamefont {A.}~\bibnamefont {Rubio-Abadal}}, \bibinfo {author}
  {\bibfnamefont {T.}~\bibnamefont {Yefsah}}, \bibinfo {author} {\bibfnamefont
  {V.}~\bibnamefont {Khemani}}, \bibinfo {author} {\bibfnamefont {D.~A.}\
  \bibnamefont {Huse}}, \bibinfo {author} {\bibfnamefont {I.}~\bibnamefont
  {Bloch}}, \ and\ \bibinfo {author} {\bibfnamefont {C.}~\bibnamefont
  {Gross}},\ }\href@noop {} {\bibfield  {journal} {\bibinfo  {journal}
  {Science}\ }\textbf {\bibinfo {volume} {352}},\ \bibinfo {pages} {1547}
  (\bibinfo {year} {2016})}\BibitemShut {NoStop}%
\bibitem [{\citenamefont {White}\ and\ \citenamefont
  {Feiguin}(2004)}]{white2004real}%
  \BibitemOpen
  \bibfield  {author} {\bibinfo {author} {\bibfnamefont {S.~R.}\ \bibnamefont
  {White}}\ and\ \bibinfo {author} {\bibfnamefont {A.~E.}\ \bibnamefont
  {Feiguin}},\ }\href@noop {} {\bibfield  {journal} {\bibinfo  {journal}
  {Physical review letters}\ }\textbf {\bibinfo {volume} {93}},\ \bibinfo
  {pages} {076401} (\bibinfo {year} {2004})}\BibitemShut {NoStop}%
\bibitem [{ITe()}]{ITensor}%
  \BibitemOpen
  \href@noop {} {\bibinfo  {journal} {\mbox{ITensor Library} (version 2.0.11)
  http://itensor.org}\ }\BibitemShut {NoStop}%
\end{thebibliography}%
\bibliographystyle{apsrev4-1}

\end{document}